\documentstyle[12pt]{article}

\begin{document}

\centerline{\bf CHAOTIC INFLATION WITH TIME-VARIABLE SPACE DIMENSIONS}
\vspace*{0.37truein}
\centerline{Forough
Nasseri$^{a,c,}$\footnote{E-mail:nasseri@theory.ipm.ac.ir},
Sohrab Rahvar$^{b,c}$}
\vspace*{0.015truein}
\centerline{$^a$\it Physics Department, Shahrood University,
P.O.Box 36155-316, Shahrood, Iran}
\baselineskip=10pt
\centerline{$^b$\it Physics Department, Sharif University of
Technology, P.O.Box 11365-9161, Tehran, Iran}
\baselineskip=10pt
\centerline{$^c$\it Institute for Studies in Theoretical
Physics and Mathematics, P.O.Box 19395-5531, Tehran, Iran}
\vspace*{0.21truein}

\begin{abstract}
Assuming the space dimension is not constant but decreases
during the expansion of the Universe, we study chaotic
inflation with the potential $m^2 \phi^2/2$. Our investigations
are based on a model Universe with variable space
dimensions. We write down field equations in the
slow-roll approximation, and define slow-roll
parameters by assuming the number of space dimensions 
decreases continuously as the Universe expands.
The dynamical character of the space dimension shifts
the initial and final value of the inflaton field to larger values.
We obtain an upper limit for the space dimension at the Planck length.
This result is in agreement with previous works for the effective time
variation of the Newtonian gravitational constant in a model Universe 
with variable space dimensions. 
\end{abstract}

\section{Introduction} \label{sec-introduction}
\noindent
In recent works \cite{11a,11b,12a}, the wave function and the dynamics of
a model Universe with variable space dimension have been studied
in detail. In this model, the number of space dimensions decreases 
continuously during the expansion of the Universe. 
There is a dimensional constraint that relates the space dimension to 
the size of the Universe. Here we apply this constraint to a simple model 
of chaotic inflation. A brief review of chaotic inflation with variable 
space dimensions has been given in \cite{capp}.

Inflationary cosmology has been studied in higher dimensional
models; for a recent work see Ref. \cite{ama}.
Cosmological scenarios of the new inflationary type form a
self-similar fractal of dimension slightly less than three \cite{mar}.
Inflation with fluctuating dimensions describes the Universe as an
exponentially large number of independent inflationary domains
(mini-Universes) of different dimension \cite{admi}.
Inflation can describe the transition from higher dimensions to the
ordinary three-space and can also be studied like a dynamical effect of
higher dimensions \cite{msz}.

In this paper, we study chaotic inflation in the framework
of model Universe with variable space dimensions.
This new investigation is useful as a conceivable application of
the idea of variable space dimensions, and as a cosmological test
of model Universe with variable space dimensions. A more appropriate  
cosmological test like nucleosynthesis or some other well established  
and precise aspect of cosmology within this speculative model are 
in progress \cite{nr}.

Our main results in this paper are: {\it i}) an upper limit for the space
dimension at the Planck length, {\it ii}) a small shift in the value of
$\phi_i$ and $\phi_f$ to larger values.
As we shall see, in this model the number of space dimensions at the
Planck length must be less than 10, which is in agreement with the
previous result in \cite{12a}.
We will use a natural unit system that sets $k_B$, $c$ and $\hbar$
all equal to $1$, so that $l_P=M_P^{-1}=\sqrt G$.

The paper is organized as follows.
In Sec. 2, we review a model Universe with variable space dimension.
In Sec. 3, chaotic inflation is formulated in an arbitrary constant space
dimension. In Sec. 4, we study chaotic inflation by assuming the space
dimension is a dynamical parameter and decreases continuously as the
Universe expands. Sec. 5 contains our conclusions.
\section{Review of a model Universe with variable space dimensions}
\subsection{Motivation for choosing a model Universe with variable
space dimensions}
\noindent
Problems of the standard model and its difficulties with the concept of
quantum gravity and the early Universe at the Planck time provide
us enough reasons to look for viable model Universes.
Moreover, the ongoing experiments related to CMB will provide
us a wealth of data suitable to test all the theories of space-time
and gravity. Even the act of verifying cosmological models based on
general relativity needs looking for viable theories differing
from it to see the degree of its testability and viability.
These are the main reasons we are studying chaotic inflation based
on time variable space dimensions. It has been shown \cite{11a}
that this idea can be implemented successfully in a gravitational
theory and cosmological model based on it. The free parameter
of the theory may then be fixed by observational data \cite{12a}.

The idea of having space-time dimensions other than $3+1$ goes
back to Kaluza-Klein theory. The generalization of this concept to
string theories with space dimension more than three, but still an
integer, and a constant is well known \cite{{tkal},{mjdu}}. This,
being considered for the high energy limit in the Universe or for
the dimension of space at the Planck time, has encouraged people
to suggest that the dimension of space in the lower energy limit,
or for the actual structured Universe, be other than three.

A possible time dependence of space-time dimensionality is derived
as an effect of entropy conservation in \cite{gaspr}. There it
turns out that, going back in time, the dimension increases first
very slowly up to about the Planck time, and increases more
rapidly thereafter. This generic trend is insensitive to the
assigned entropy value. In fact a minimum value for the size of
the Universe, being about the Planckian size, is also obtained.
The dynamical model we are going to follow has the same time
behavior for the space dimension. A minimum size for the Universe
is also built into our dynamical model.

Our treatment in this paper is based on a cosmological model, 
where the number of spatial dimensions decreases continuously as the
Universe expands, presented in the pioneer works \cite{{11a},{9a},{10a}}.
A proposed way of handling such a concept is using the idea of decrumpling
coming from polymer physics \cite{{dnel},{ffab},{pjde}}.
In this model the fundamental building blocks of
the Universe are like cells being arbitrary dimensions and having, in
each dimension, a characteristic size $\delta$ which maybe of the order of
the Planck length ${\mathcal O}$($10^{-33}$ cm) or even smaller.
These ``space cells'' are embedded in a ${\mathcal D}$ space, where
${\mathcal D}$ may be up to infinity. 
Therefore, the space dimension of the
Universe depends on how these fundamental cells are configured in this
embedding space. The Universe may have begun from a very crumpled
state having a very high dimension ${\mathcal{D}}$ and a size $\delta$,
then have lost dimension through a uniform decrumpling which we see
like a uniform expansion. The expansion of space, being now understood like
a decrumpling of cosmic space, reduces the space-time dimension continuously
from ${\mathcal{D}}+1$ to the present value $D_0+1$.
In this picture, the Universe can have any space dimension.
As it expands, the number of spatial dimensions decreases continuously.
The physical process that causes or necessitates such a decrease in
the number of spatial dimensions comes from how these fundamental cells
are embedded in a ${\mathcal{D}}$ space.

As an example, take a limited number of small three-dimensional beads.
Depending on how these beads are embedded in space they can configure to a
one-dimensional string, two-dimensional sheet, or three-dimensional sphere.
This is the picture we are familiar with from the concept of crumpling
in polymer physics where a crumpled polymer has a dimension more than 1.
Or take the picture of a clay which can be like a
three-dimensional sphere, or a two-dimensional sheet, or even a
one-dimensional string, a picture based on the theory of fluid membranes.

The major formal difficulty to implement this idea in a
space-time theory with variable space dimension is that the
measure of the integral of the action is variable and therefore
some part of integrand. However, taking into account the
cosmological principle, i.e. the homogeneity and isotropy of
space, the formulations are simplified substantially. It then becomes
possible to formulate a Lagrangian for the theory and write down the
corresponding field equations. This Lagrangian is however 
not unique \cite{11a}.

The original decrumpling model of the Universe seems to be
singularity free, having two turning points for the space
dimension \cite{9a}. Lima and Mohazzab \cite{10a}
criticize the way of generalizing the standard cosmological model
to arbitrary variable space dimension used in \cite{9a} and propose another
way of writing the field equations. Their model shows no upper
bound for the dimension of space. Later on this scenario was
extended to the class of multidimensional cosmological models,
where extra factor spaces play the role of the matter fields. In
this multidimensional cosmological model an inflationary solution
was found together with the prediction that the Universe starts
from a nonsingular space time \cite{uble}.

A new way to generalize the gravitational action in constant
dimension to the case of dynamical dimension is recently
proposed by Mansouri and Nasseri \cite{11a}. 
There, it is shown that the generalization of the gravitational 
action to the dynamical
dimension is not unique. Moreover, in contrast to the earlier
works in \cite{{9a},{10a}}, the dependence of the measure of
the action on space dimension is taken into account.
This new decrumpling model is studied in detail \cite{11a}.
The generalization of the action, the Lagrangian, the equations of
motion to dynamical space dimension, the time evolution of the
spatial dimension, numerical results for the turning points of the model,
and its quantum cosmology within the concept of the Wheeler-DeWitt
equation are derived. It is shown that the corresponding potential of
the model has completely different behavior from the potential of the
de Sitter minisuperspace in three-space. Imposing the appropriate boundary
condition in the limit $ a \to +\infty$, and using the
semiclassical approximation, the wave function of the model is
also derived. It is then seen that in the limit of constant space
dimension, the wave function is not well-defined. It can approach
to the Hartle-Hawking wave function or to the modified Linde wave
function, but not to that of Vilenkin. In the limit of constant
spatial dimension, the probability density approaches to Vilenkin,
Linde and others' proposal; i.e. to the probability
density ${\mathcal{P}} \propto \exp(2S_E)$, or more generally
$\exp(-2|S_E|)$, where $S_E$ is the Euclidean action of the
classical instanton solution.

\subsection{Relation between the effective space dimension
$D(t)$ and characteristic size of the Universe $a(t)$}
\noindent
Assume the Universe consists of a fixed number $N$ of universal
cells having a characteristic length $\delta$ in each of their
dimensions. The volume of the Universe at the time $t$ depends
on the configuration of the cells. It is easily seen that
\cite{{9a},{10a}}
\begin{equation}
{\rm vol}_D({\rm cell})={\rm vol}_{D_0}({\rm cell})\delta^{D-D_0}.
\end{equation}

Interpreting the radius of the Universe, $a$, as the radius of
gyration of a crumpled ``universal surface''\cite{pjde},
the volume of space can be written \cite{{9a},{10a}}
\begin{eqnarray}
a^D&=&N {\rm vol}_D({\rm cell})\nonumber\\
   &=&N {\rm vol}_{D_0}({\rm cell}) \delta^{D-D_0}\nonumber\\
   &=&{a_0}^{D_0} \delta^{D-D_0}
\end{eqnarray}
or
\begin{equation}
\label{31}
\left( \frac{a}{\delta} \right)^D=
\left( \frac{a_0}{\delta} \right)^{D_0} = e^C,
\end{equation}
where $C$ is a universal positive constant. Its value has a strong
influence on the dynamics of space-time, for example on the dimension
of space, say, at the Planck time. Hence, it has physical and cosmological
consequences and may be determined by observation. The zero subscript in any
quantity, e.g. in $a_0$ and $D_0$, denotes its present values.
We coin the above relation as a``dimensional constraint" which relates
the ``scale factor" of our model Universe to the space dimension.
In our formulation, we consider the comoving length of the Hubble radius
at present time to be equal to one. So the interpretation of the scale
factor as a physical length is valid.
The dimensional constraint can be written in this form
\begin{equation}
\label{32}
\frac{1}{D}=\frac{1}{C}\ln \left( \frac{a}{a_0} \right) + \frac{1}{D_0}.
\end{equation}

It is seen that by expansion of the Universe, the space
dimension decreases. Note that in Eqs.(\ref{31}) and (\ref{32}),
the space dimension is a function of cosmic time $t$.
Time derivative of Eqs.(\ref{31}) or (\ref{32}) leads to

\begin{equation}
\label{33}
\dot{D}=-\frac{D^2 \dot{a}}{Ca}.
\end{equation}
It can be easily shown that the case of constant space dimension
corresponds to when $C$ tends to infinity. In other words,
$C$ depends on the number of fundamental cells. For $C \to \infty$,
the number of cells tends to infinity and $\delta\to 0$.
In this limit, the dependence between the space dimensions and
the radius of the Universe is removed, and consequently we
have a constant space dimension.

\subsection{Lagrangian formulations of the model and field equations}
\noindent
Usually, we are accustomed to work with an integer number of
dimension, and therefore a non-integer total number of
spatial dimensions looks peculiar.
It is clear that for non-integer value of space dimensions, one cannot
define the metric tensor. To overcome this problem, we use a
gravitational theory based on Lagrangian formulations.
In Ref. \cite{11a}, some shortcomings of the original Lagrangian
formulation of the model proposed in \cite{9a} have been
shown, regarding the fields equations and their results.

Let us define the action of the model for the special
Friedmann-Robertson-Walker (FRW) metric in an arbitrary fixed space
dimension $D$, and then try to generalize it to variable dimension.
Now, take the metric in constant $D+1$ dimensions in the following
form
\begin{equation}
ds^2 = -N^2(t)dt^2+a^2(t)d\Sigma_k^2,
\end{equation}
where $N(t)$ denotes the lapse function and $d\Sigma_k^2$ is the line
element for a D-manifold of constant curvature $k = + 1, 0, - 1$. The
Ricci scalar is given by
\begin{equation}
\label{r}
R=\frac{D}{N^2}\left\{\frac{2\ddot a}{a}+(D-1)\left[\left(\frac{\dot a}{a}
\right)^2 + \frac{N^2k}{a^2}\right]-\frac{2\dot a\dot N}{aN}\right\}.
\end{equation}
Substituting from Eq.(\ref{r}) in the Einstein-Hilbert action for
pure gravity,
\begin{equation}
\label{s}
S_G = \frac{1}{2\kappa} \int d^{(1+D)} x \sqrt{-g}R,
\end{equation}
and using the Hawking-Ellis action (see Ref. [1]) of a perfect fluid,
for the model Universe with variable space dimension, the following
Lagrangian has been obtained \cite{11a}
\begin{eqnarray}
\label{34}
L_I&:=& -\frac{V_D}{2 \kappa N} \left( \frac{a}{a_0} \right)
^D D(D-1)
\left[ \left( \frac{\dot a}{a} \right )^2 -\frac{N^2 k}{a^2} \right ]\nonumber\\
&-& \rho N V_D \left( \frac{a}{a_0} \right )^D,
\end{eqnarray}
where $\kappa=8 \pi {M_P}^{-2}$, $\rho$ the energy density, and $V_D$
the volume of the space-like sections
\begin{eqnarray}
\label{36}
V_D &=& \cases {\frac{2 \pi^{(D+1)/2}}{\Gamma[(D+1)/2]},
	       & if $\;k=+1$, \cr
	       \frac{\pi^{(D/2)}}{\Gamma(D/2+1)}{\chi_c}^D,
	       & if $\;k=0$, \cr
	       \frac{2\pi^{(D/2)}}{\Gamma(D/2)}f(\chi_c),
	       & if $\;k=-1$. \cr}
\end{eqnarray}
Here $\chi_C$ is a cut-off and $f(\chi_c)$ is a function thereof
(see Ref. \cite{11a}).
In the limit of constant space dimensions, or $D=D_0$,
$L_I$ approaches to the Einstein-Hilbert Lagrangian
which is
\begin{eqnarray}
\label{37}
L_{I}^0 &:=& - \frac{V_{D_0}}{2 \kappa N}
\left( \frac{a}{a_0} \right)^{D_0} D_0(D_0-1)
\left[ \left( \frac{\dot{a}}{a} \right)^2 - \frac{N^2 k}{a^2} \right ]\nonumber\\
&-& \rho N V_{D_0} \left( \frac{a}{a_0} \right )^{D_0}.
\end{eqnarray}
So, Lagrangian $L_I$ cannot abandon Einstein's gravity.
Varying the Lagrangian $L_I$ with respect to $N$ and $a$, we find the
following equations of motion in the gauge $N=1$, respectively
\begin{eqnarray}
\label{38}
&&\left( \frac{\dot a}{a} \right)^2 +\frac{k}{a^2} =
\frac{2 \kappa \rho}{D(D-1)},\\
\label{39}
&&(D-1) \left\{ \frac{\ddot{a}}{a} + \left[ \left( \frac{\dot a}{a}
\right)^2
+\frac {k}{a^2} \right] \left( -\frac{D^2}{2C} \frac{d \ln V_D}{dD}-
1-\frac{D(2D-1)}{2C(D-1)}+\frac{D^2}{2D_0} \right) \right\}\nonumber\\
&&+ \kappa p \left( -\frac{d \ln V_D}{dD} \frac{D}{C} - \frac{D}{C}
\ln \frac{a}{a_0} +1 \right) =0.
\end{eqnarray}
Using (\ref{33}) and (\ref{38}), the evolution equation of the space
dimension can be obtained by
\begin{equation}
\label{310}
{\dot{D}}^2= \frac{D^4}{C^2} \left[ \frac{2 \kappa \rho}{D(D-1)}
-k {\delta}^{-2} e^{-2C/D} \right].
\end{equation}
The continuity equation of the model Universe with variable space
dimension
can be obtained by (\ref{38}) and (\ref{39})
\begin{equation}
\label{311}
\frac{d}{dt} \left[ \rho \left( \frac{a}{a_0} \right)^D V_D \right]
+ p \frac{d}{dt} \left[ \left( \frac{a}{a_0} \right )^D V_D \right] =0.
\end{equation}
This continuity equation can be integrated for the case of dust or radiation
to obtain the energy density like a function of time. For the radiation era,
$p=\rho/D$, we have
\begin{equation}
\label{312}
\rho = \rho_{eq} e^{C(D/D_{eq} -1)} \left( \frac{D}{D_{eq}}
\right)^{C/D_{eq}}
\frac{V_{D_{eq}}}{V_D} \exp \left(- \int^D_{D_{eq}} dD \frac{1}{D}
\frac{d \ln V_D}{dD} \right),
\end{equation}
and for the matter era, $p=0$,
\begin{equation}
\label{313}
\rho=\rho_0 e ^{C(D/D_0 -1)} \frac{V_{D_0}}{V_D},
\end{equation}
where $D_{eq}$ is the space dimension of the Universe when the scale
factor is equal to the scale factor of the equality
epoch \footnote{Note that Eq. (70) of Ref. \cite{11a}
must be corrected by replacing $D_{eq}$ and $\rho_{eq}$
with $D_0$ and $\rho_0$, respectively. For the radiation
era, we cannot use the present values of the energy density $\rho_0$
and the space dimension $D_0$. For the matter era, Eq. (71)
of Ref. \cite{11a} is correct.}.

We define $D_P$ like the space dimension of the Universe when the
scale factor is equal to the Planck length $l_P$.
Taking the scale of the Universe at $D_0=3$ to be the present value of the
Hubble radius $H_0^{-1}$ and the space dimension in the Planck length to be
$4, 10,$ or $25$, from Kaluza-Klein and superstring theories, we can
obtain from Eqs. (\ref{31}) and (\ref{32}) the corresponding value of
$C$ and $\delta$. In Table 1, values of $C$ and $\delta$ for some
interesting values of $D_P$ are given. These values are calculated by
assuming $D_0=3$ and
$H_0^{-1}=3000 {h_0}^{-1} {\rm Mpc} = 9.2503 \times 10^{27} {h_0}^{-1} 
{\rm cm}$. Since the value of $C$ and $\delta$ are not very sensitive to
$h_0$ we take $h_0=1$.
\begin{table}
\begin{center}
\caption{Values of $D_P$, $C$, $\log(\delta/l_P)$, $\delta$,
$D_f$, $D_{eq}$, $|1-\Omega^{-1}_f|$.
We take $D_0=3$ and $a_0=H_0^{-1}= 9.2503 \times 10^{27} {\rm cm}$.} 
\begin{tabular}{cccccccc} \hline
$D_P$ & $C$ & $\log_{10}(\delta/l_P)$ & $\delta ({\rm cm})$ & $D_f$ &
$D_{eq}$ & $|1-\Omega^{-1}_f|$ \\ \hline
$3$ & $+\infty$ & $-\infty$ & $0$ & $3.00$ & $3.00$ &
$2.190 \times 10^{-51}$ \\
$4$  & $1678.8$ & $-182.270$ & $8.6 {\times}10^{-216}$ & $3.385$ &
$3.058$ & $1.628 \times 10^{-61}$ \\
$10$ & $599.57$ & $-26.03$ & $1.5 \times 10^{-59}$ &
$4.401$ & $3.168$ & $6.577 \times 10^{-88}$ \\
$25$  &  $476.93$   & $-8.28$    & $8.4 \times10^{-42}$ & $5.001$ &
$3.214$ & $2.783 \times 10^{-103}$ \\
$+\infty$ &  $419.70$  & $0$  & $l_P$ & $5.501$ & $3.245$ &
$5.227 \times 10^{-116}$ \\ \hline
\end{tabular}
\end{center}
\end{table}

\section{Inflationary cosmology with constant space dimensions}
Assuming the space dimension is constant and has an arbitrary value,
we now formulate the inflationary cosmology in a constant D-space.
The crucial equations are given by \cite{{11a},{12a}}
\begin{eqnarray}
\label{41}
&&H^2=\frac{16 \pi \rho}{D(D-1)M_P^2}-\frac{k}{a^2}\;\;\;\;\;
{\mbox{Friedmann equation}},\\
\label{42}
&&\dot{\rho}+DH(\rho+p)=0\;\;\;\;\;\;\;\;\;\;\;{\rm {Fluid\;\;equation}}.
\end{eqnarray}
Inserting $k=0$ and the energy density and pressure of a homogeneous
inflaton field
\begin{eqnarray}
\label{42a}
\rho &\equiv& \frac{1}{2} {\dot \phi}^2 +V(\phi),\\
\label{42b}
p &\equiv& \frac{1}{2} {\dot \phi}^2 - V(\phi),
\end{eqnarray}
in Eqs. (\ref{41}) and (\ref{42}), we are led to
\begin{eqnarray}
\label{43}
&&H^2 = \frac{16 \pi}{D(D-1) M_P^2} \left( \frac{1}{2} {\dot \phi}^2+
V(\phi) \right),\\
\label{44}
&&\ddot{\phi}+DH \dot{\phi} = - V'(\phi).
\end{eqnarray}
In a constant D-space the slow-roll conditions are defined by
\begin{equation}
\label{45}
{\dot \phi}^2 \ll V(\phi),\;\;\;\ddot\phi \ll DH \dot{\phi},\;\;\;
-\dot H \ll H^2.
\end{equation}
Using these conditions, Eq. (\ref{43}) and (\ref{44}) can be
written in the simpler set
\begin{eqnarray}
\label{46}
H^2&=& \frac{16 \pi V(\phi)}{D(D-1) M_P^2},\\
\label{47}
DH \dot{\phi} &\simeq& -V'(\phi).
\end{eqnarray}
During inflation, $H$ is slowly varying in the sense that its change per
Hubble time, $\epsilon \equiv -{\dot H}/H^2$ is less than one
\cite{{arli},{34a}}. The slow-roll condition $|\eta| \ll 1$ is
actually a consequence of the condition $\epsilon \ll 1$ plus the
slow-roll
approximation $D H \dot\phi \simeq -V'$. Indeed, differentiating
Eq. (\ref{47}) one finds
\begin{equation}
\label{49}
\frac{\ddot \phi}{H \dot \phi}= \epsilon - \eta,
\end{equation}
where the slow-roll parameters in any
constant space dimension are defined by
\begin{eqnarray}
\label{410}
&&\epsilon \equiv -\frac{\dot H}{H^2} =\frac{(D-1)M_P^2}{32 \pi}
\left( \frac{V'}{V} \right)^2,\\
\label{411}
&&\eta \equiv \frac{(D-1)M_P^2}{16 \pi} \left( \frac{V''}{V} \right).
\end{eqnarray}
The number of e-foldings between $t_i$ and $t_f$ is given by
\begin{equation}
\label{412}
{\mathcal{N}}=\int^{t_f}_{t_i} H(t) dt = 
\ln \left( \frac{a_f}{a_i}\right)
\simeq -\frac{16 \pi}{(D-1) M_P^2} \int^{\phi_f}_{\phi_i}
\frac{V}{V'} d\,\phi.
\end{equation}
For the chaotic inflation with the $m^2 \phi^2/2$ potential, the
solution of Eqs. (\ref{46}) and (\ref{47}) are given by
\begin{eqnarray}
\label{413}
\phi(t)&=&\phi_i-\frac{m M_P}{2} \sqrt{\frac{D-1}{2 \pi D}}\,t,\\
\label{414}
a(t)&=& a_i \exp \left( \frac{4 \pi}{M_P^2(D-1)}
\left[ \phi_i^2-\phi^2(t) \right] \right).
\end{eqnarray}
Using the slow-roll parameters
\begin{equation}
\label{415}
\epsilon=\eta=\frac{M_P^2 (D-1)}{8 \pi \phi^2},
\end{equation}
and the failure of the slow-roll conditions
\begin{equation}
\label{416}
{\rm max}\left\{ \epsilon_f; |\eta_f| \right\} \simeq 1,
\end{equation}
one concludes that
\begin{equation}
\label{fif}
\phi_f=M_P \sqrt{\frac{D-1}{8 \pi}}.
\end{equation}
Substituting this value of $\phi_f$ into Eq. (\ref{412}),
one can get
\begin{equation}
\label{417}
{\mathcal{N}}=\frac{4 \pi \phi_i^2}{(D-1) M_P^2} - \frac{1}{2}.
\end{equation}
Considering the upper bound of the number of e-foldings, ${\mathcal{N}} >
60$, we obtain
\begin{equation}
\label{418}
\phi_i > \frac{1}{2} \sqrt{\frac{60}{\pi}(D-1)} M_P.
\end{equation}
From Eq. (\ref{414}), we see that by the time $\phi$ approaches zero
the Universe has expanded by an inflating factor
\begin{equation}
\label{419}
l \sim l_P \exp \left( \frac{4 \pi \phi_i^2}{M_P^2(D-1)} \right)
\sim l_P \exp \left( \frac{4 \pi M_P^2}{m^2 (D-1)} \right),
\end{equation}
where we use the condition $\phi_i \leq {M_P^2}/{m}$.
For $m \sim 10^{-6} M_P$, this implies that the inflationary
domains of the Universe expand to
\begin{equation}
\label{420}
l \sim 10^{-33} \exp \left( \frac{4 \pi \times 10^{12}}{D-1}
\right).
\end{equation}
For ${1< D} \leq 6$ we have $l > 10^{10^{12}}$.
In the limit of three-space, from Eqs. (\ref{fif}, \ref{417},
\ref{418}, \ref{419}), we obtain:
\begin{eqnarray}
\label{421b1}
&&\phi_f=M_P/{\sqrt{4\pi}},\\
\label{421b2}
&&\phi_i > \sqrt{\frac{30}{\pi}}M_P\approx 3.10 M_P,\\
\label{421b3}
&&{\mathcal{N}} \approx 60,\\
\label{421b4}
&&l \sim 10^{-33} \exp (2 \pi 10^{12}) > {10^{10}}^{12} {\rm {cm}}.
\end{eqnarray}
Inflation ends when $t=t_f$. Using Eq.(\ref{413}), one can obtain:
\begin{equation}
\label{421a}
t_f=\frac{2 \sqrt{2 \pi D}(\phi_i-\phi_f)}{mM_P \sqrt{D-1}}.
\end{equation}
For $D=3$, we substitute the value of $\phi_i/M_P$ and $\phi_f/M_P$ from
Eqs. (\ref{421b1},\ref{421b2}) in Eq.(\ref{421a}). Therefore, we obtain:
\begin{equation}
\label{421}
t_f \simeq 7.71 \times 10^{-37} {\rm {sec}},
\end{equation}
where we take $m \simeq 1.21 \times 10^{-6} M_P$, see \cite{{8a8},{rbra}}.
\section{Chaotic inflation with time-variable space dimensions}
Let us now study the inflationary cosmology when the space dimension is
a dynamical parameter. To do this, we first write down our formulation
for a general potential and then use our results in the $m^2\phi^2/2$
potential.
\subsection{General formulations}
To study inflation in the framework of the model Universe with variable
space dimension, the crucial equations are obtained by substituting
Eqs. (\ref{42a}) and (\ref{42b}) into Eqs. (\ref{38}) and (\ref{311}).
Therefore, one can find
\begin{eqnarray}
\label{51}
&&H^2=\frac{16 \pi}{D(D-1)M_P^2} \left( \frac{1}{2} {\dot\phi}^2 +
V(\phi) \right) -\frac{k}{a^2}\;\;\;\;\;{\mbox{Friedmann equation}},\\
\label{52}
&&\ddot{\phi}+DH\dot\phi+\dot{D} \dot\phi \left( \ln \frac{a}{a_0} +
\frac{d \ln V_D}{dD} \right) = -V'(\phi)\;\;\;\;\;
{\mbox{Fluid equation}}.
\end{eqnarray}
In the case of dynamical space dimension, the slow-roll conditions are the
same as in that of constant space dimension, see Eq. (\ref{45}).
It is worth mentioning that in the slow-roll condition
$\ddot\phi \ll DH\dot\phi$, where $D$ is a dynamical parameter.
When inflation starts, the curvature term can be neglected, $k=0$.
So, in the slow-roll approximation, Eqs. (\ref{51}) and (\ref{52})
lead to
\begin{eqnarray}
\label{53}
&&H^2\simeq \frac{16 \pi V(\phi)}{D(D-1) M_P^2},\\
\label{54}
&& DH \dot\phi +\dot{D} \dot{\phi} \left( \ln \frac{a}{a_0}+
\frac{d \ln V_D}{d D} \right) \simeq - V'(\phi),
\end{eqnarray}
where $V_D$ is the space-like section for $k=0$ and 
\begin{equation}
\label{55}
\frac{d \ln V_D}{d D} = \ln \chi_c + \frac{1}{2} \ln \pi -
\frac{1}{2} \psi \left( \frac{D}{2}+1 \right).
\end{equation}
Here Euler's psi function $\psi$ is the logarithmic derivative of the
gamma function, $\psi (x) \equiv \Gamma'(x)/\Gamma (x)$.
Substituting the dimensional constant (\ref{33}) into Eq. (\ref{53}),
the dynamics of the space dimension is given by
\begin{equation}
\label{56}
{\dot{D}}^2 \simeq \frac{16 \pi D^3 V(\phi)}{C^2 (D-1) M_P^2}.
\end{equation}
Using Eqs. (\ref{53}) and (\ref{54}), one can find
\begin{equation}
\label{57}
\frac{\ddot\phi}{H \dot\phi}= \frac{\epsilon}{D \left( \frac{1}{D_0}
-\frac{1}{C} \frac{d \ln V_D}{dD} \right) }-\eta,
\end{equation}
where the slow-roll parameters in the variable space dimension are defined
by
\begin{eqnarray}
\label{58}
\epsilon&\equiv& - \frac{\dot H}{H^2} = \frac{(D-1)M_P^2}
{32 \pi D \left( \frac{1}{D_0} - \frac{1}{C} \frac{d\ln V_D}{dD} \right) }
\left( \frac{V'}{V} \right)^2 + \frac{D(1-2D)}{2C (D-1)},\\
\label{59}
\eta &\equiv& \frac{(D-1) M_P^2}{16 \pi D \left( \frac{1}{D_0} -
\frac{1}{C} \frac{d \ln V_D}{dD} \right) } \left( \frac{V''}{V} 
\right)\nonumber\\
&+&\frac{\frac{-2}{C}+\frac{D^2}{C^2} \frac{d^2\ln V_D}{dD^2}
+\frac{\ddot D}{H^2 D^2} \left( \ln \frac{a}{a_0} +\frac{d \ln V_D}{dD}
\right)}
{D \left( \frac{1}{D_0} - \frac{1}{C} \frac{d \ln V_D}{d D} \right)}.
\end{eqnarray}
Note that in the limit of $C \to +\infty$ and $D=D_0$, 
these slow-roll parameters approach Eqs. (\ref{410}) and (\ref{411}),
respectively.

One can get the critical
density\footnote{Substituting $k=0$ and $\rho=\rho_C$ in Eq. (\ref{41}),
one gets the same expression for $\rho_C$ as when the space dimensions are
constant. So, the definition of $\rho_C$ is the same regardless of whether
the space dimensions are constant.} by taking $k=0$ in Eq. (\ref{38})
\begin{equation}
\label{62}
\rho_C \equiv \frac{H^2 D (D-1) M_P^2}{16 \pi}.
\end{equation}
From Eqs. (\ref{38}) and (\ref{62}), the present value of $\Omega$
can be obtained by
\begin{equation}
\label{63}
\Omega_0=1+\frac{k}{a_0^2 H_0^2}.
\end{equation}
Here our treatments are based on recent observational data
$\Omega_0=1$. Substituting
$k=0$ and $\Omega_0=1$ into Eq. (\ref{63}), we obtain
\begin{equation}
\label{64}
a_0=H_0^{-1}=9.2503 \times 10^{27} {\rm cm}.
\end{equation}
Using Eqs. (\ref{31}) and (\ref{64}), one can find the value of $C$ and
$\delta$ for $D_P=4, 10, 25,$ and $+\infty$ with $D_0=3$, see Table 1.

\subsection{Chaotic inflation with variable space dimensions}
\noindent
In variable space dimensions, the general potential for chaotic
inflation can be taken as $V \propto \phi^{\alpha}$. In higher
dimensions, the value of $\alpha$ should be specified so that the
theory could be renormalizable. We are interested now in studying the
simplest chaotic inflation with the $m^2 \phi^2/2$ potential in
the model Universe with variable space dimensions. We begin by
calculating the number of e-foldings. To do this, one needs the
value of $|1-\Omega^{-1}_f|$. Using Eqs. (\ref{38}) and
(\ref{62}), we are led to
\begin{equation}
\label{65}
|1-\Omega^{-1}_f| = \frac{D_f (D_f -1) M^2_P |k|} {16 \pi \rho_f a^2_f},
\end{equation}
where the subscript $f$ denotes the value at the GUT epoch, i.e.
$\rho_f \equiv \rho_{GUT}$, $a_f \equiv a_{GUT}$,
$D_f \equiv D_{GUT}$ and $|1-\Omega^{-1}_f| \equiv |1-\Omega^{-1}_{GUT}|$.
We are going to obtain the value of $|1-\Omega^{-1}_f|$ in Eq. (\ref{65}).
To do this, we must know $a_f$, $D_f$, and $\rho_f$.
To obtain the value of $a_f$, we start from the equality time, go back 
in time, and follow the standard cosmology.
At some time $t=t_{eq}$ in the past (corresponding to a value
$a=a_{eq}$ and redshift $z=z_{eq}$ ) the radiation and matter will
have equal density and we have \cite{4a}
\begin{equation}
\label{66}
(1+z_{eq})=\frac{a_0}{a_{eq}} \simeq 3.9 \times 10^{4}
({\rm {\Omega}}\,{h^2}).
\end{equation}
Once the temperature of the radiation grows as $a^{-1}$, the temperature
of the Universe at this epoch will be
\begin{equation}
\label{67}
T_{eq}=T_{now} (1+z_{eq})=9.24\,({\rm \Omega}\,h^2)\,{\rm eV}.
\end{equation}
Using $T_f \simeq 10^{24}\,({\rm \Omega}\,h^2)\,{\rm eV}$ and thanks to
the fact that $(a_f/a_{eq})=(T_{eq}/T_f)$, we obtain
$a_f \simeq 2.19 {\rm cm}$.
Substituting this value into the dimensional constraint (\ref{32}),
the corresponding value of $D_f$ can be obtained. The space dimension
$D=D_{eq}$, when $a=a_{eq}$, can be obtained by Eqs. (\ref{32}) and
(\ref{66}), see Table 1.
From Eq. (\ref{313}), the energy density at the time $t=t_{eq}$ is given
by
\begin{equation}
\label{68}
\rho_{eq}=\rho_0 e ^{C(D_{eq}/D_0 -1)} \frac{V_{D_0}}{V_{D_{eq}}}.
\end{equation}
Substituting this into Eq. (\ref{312}), one can get the energy density at
the GUT epoch
\begin{eqnarray}
\label{69}
\rho_f &=&\rho_0 e ^{C(D_{eq}/D_0 -1)} \frac{V_{D_0}}{V_{D_{eq}}}
e^{C({D_f}/D_{eq} -1)} \left( \frac{D_f}{D_{eq}}\right)^{C/D_{eq}}
\frac{V_{D_{eq}}}{V_{D_f}} \nonumber\\
&\times& \exp \left(-\int^{D_f}_{D_{eq}} dD \frac{1}{D}
\frac{d \ln V_D}{dD} \right).
\end{eqnarray}
Based on recent observational evidence where $\Omega_0=1$ to high
accuracy, we have
$$
\rho_0 \equiv \Omega_0\,\rho_c =
1.88 \times 10^{-29} {h_0}^2\,{\mbox {gr cm}}^{-3}=
8.10 \times 10^{-47} {h_0}^2 {\rm GeV}^4.
$$
In Eq. (\ref{69}), it should be noted that the value of space-like
sections is for $k=0$. So, from Eq. (\ref{36}) one gets
\begin{equation}
\label{70}
\int^{D_f}_{D_{eq}}dD \frac{1}{D} \frac{d \ln V_D}{dD} =
\left( \ln \chi_c + \frac{\ln \pi}{2} \right) \ln \frac{D_f}{D_{eq}} -
\frac{1}{2} \int^{D_f}_{D_{eq}} dD \frac{1}{D}\psi\left( \frac{D}{2} +1 
\right),
\end{equation}
From Eqs. (\ref{65}), (\ref{69}) and (\ref{70}), one finds the value of
$|1-\Omega^{-1}_f|$,
see Table 1 \footnote{It should be noted that in the limit
of constant space dimension, we have
$$\lim_{C \to +\infty} \left( \frac{D}{D_{eq}} \right)^{C/D_{eq}}=
\lim_{C \to +\infty} \left( \frac{D}{C} \ln \frac{a_{eq}}{a}+1 \right)
^{C/D_{eq}}=\frac{a_{eq}}{a},$$
and also
$$\lim_{C \to +\infty} e^{C(D/D_{eq}-1)}=
\left( \frac{a_{eq}}{a} \right)^{D_0}.$$
So, from Eqs. (\ref{65}) and (\ref{69}), one gets in the constant
three-space
$$|1-\Omega^{-1}_f|=\frac{3 M^2_P}
{8 \pi \rho_{eq} \left( {a_{eq}}/{a_f} \right)^4 a_f^2} \simeq
2.190 \times 10^{-51}.$$}.

To calculate the number of e-foldings ${\mathcal {N}}\equiv\ln(a_f/a_i)$,
we use Eq. (\ref{65}) for some finite time interval
$t \in [t_i,t_f]$ of the inflationary epoch
\begin{equation}
\label{71}
\frac{\rho_i a_i^2 |1-\Omega^{-1}_i|}{D_i (D_i-1)}=
\frac{\rho_f a_f^2 |1- \Omega^{-1}_f|}{D_f(D_f-1)}.
\end{equation}

The inflaton is like a little ball that
rolls down the potential hill. Near the top of the potential the
slope is very small, so the roll is slow and $V(\phi) \simeq {\rm
const}$. In other words $\rho_i \simeq \rho_f$.
Considering $\rho_i \simeq \rho_f$ and
$\Omega_i \approx {\mathcal{O}}(1)$ and using
the dimensional constraint (\ref{32}), one can write down
Eq. (\ref{71}) as
\begin{equation}
\label{72}
\left( \frac{a_f}{a_i} \right)^2 \left(
\frac{1}{\frac{1}{C}\ln \frac{a_f}{a_i} - \frac{1}{D_f}} \right)
\left( \frac{1}{\frac{1}{C} \ln \frac{a_f}{a_i} - \frac{1}{D_f}}+1
\right) = \frac{D_f (D_f -1)}{|1-\Omega^{-1}_f|}.
\end{equation}
From this equation, we find the value of ${\mathcal{N}}$ by numerical
calculations.
The initial size of the Universe $a_i$ at the beginning of inflation
is given by $a_i=a_f \exp(-{\mathcal{N}})$.
Using the number of e-folding and Eq. (\ref{32}), it is easy to calculate
the space dimension $D_i$ at the beginning of inflation.
The values of $\mathcal{N}$, $D_i$ and $a_i$ are given in Table 2.

\begin{table}
\begin{center}
\caption{Values of $\mathcal{N}$, $D_i$, $a_i$, $\phi_f/M_P$ and
$\phi_i/M_P$. We consider the value of $D_P$, $C$, and $\delta$ as
given in Table 1. Since for $D_P \geq 10$ the value of $D_i$ is negative, 
the values of $\phi_f/M_P$ and $\phi_i/M_P$ are not calculated 
and are denoted by a star.}
\begin{tabular}{ccccccc} \hline
$D_P$ & $C$ & $\mathcal{N}$ & $D_i$ & $a_i({\rm cm})$
& $\phi_f/M_P$ & $\phi_i/M_P$ \\ \hline
$3$ & $+\infty$ &  $58.32$ & $3.00$ & $1.025 \times 10^{-25}$ &
$1/{\sqrt{4 \pi}}$ & $3.060$ \\
$4$  & $1678.8$ & $69.80$ & $3.939$ &
$1.063 \times 10^{-30}$ & $0.290$ & $3.491$ \\
$10$ & $599.57$ & $98.97$ & $16.084$ &
$2.282 \times 10^{-43}$ & $0.304$ & $4.512$ \\
$25$  &  $476.93$  & $116.42$ & $-22.653$ &
$6.024 \times 10^{-51}$ & $\star$ & $\star$ \\
$+\infty$ &  $419.70$  & $132.25$ & $-7.500$ &
$8.035 \times 10^{-58}$ & $\star$ & $\star$ \\ \hline
\end{tabular}
\end{center}
\end{table}

It is worth mentioning that for $D_P \geq 25$, $D_i$ takes a negative
value. Also for $D_P=10$, $D_i$ exceeds the value of $D_P$;
in other words $a_i$ is smaller than the Planck length.
This means that the inflationary period of the history of the Universe
with variable space dimensions with $D_P=10$ takes place before the
Planck epoch. This is even more
surprising when one realizes that during inflation a region of initial
size is $\Delta l \sim l_P\sim 10^{-33} {\rm cm}$.
Therefore we rule out the cases with $D_P \geq 10$ in the model Universe
with variable space dimensions. In what follows, we will be particularly
interested in the case $D_P=4$ and $C=1678.8$ for which inflation
begins from an initial size of about
$a_i \sim 10^{-30} {\rm cm} \sim 10^3 \times l_P$ and with an initial
space dimension $D_i \simeq 3.94$.

Let us now describe how to obtain the initial and the final value of the
inflaton field denoted by $\phi_i$ and $\phi_f$, respectively.
The end of inflation is marked by the failure of one of three slow-roll
conditions, see Eq. (\ref{416}). Taking $\epsilon_f=\eta_f=1$,
one can use Eqs. (\ref{58}) and (\ref{59}) to calculate
$\phi_f/M_P$
\begin{equation}
\label{73}
\frac{\phi_f}{M_P}=\sqrt {\frac{D_f-1}{8 \pi D_f \left[ \frac{1}{D_0} -
\frac{1}{C} \ln \chi_c - \frac{1}{2C} \ln \pi +
\frac{1}{2C} \psi \left( \frac{D_f}{2} +1 \right) \right] }}.
\end{equation}

It should be noted that we neglect the second term on the RHS of
Eqs. (\ref{58}) and (\ref{59}), because these terms are independent
of the shape of the inflaton potential and make a small contribution to
the value of $\epsilon$ and $\eta$. Using Eq. (\ref{73}), one
can find $\phi_f/M_P=0.290$ for $D_P=4$, see Table 2.
Notice the value of $\phi_f/M_P$
depends on $\chi_C$. We take $\chi_c=1$
in the comoving coordinate which is about Planck length in the 
physical coordinate at the Planck time.
For more details about the value of $\chi_C$, see Ref. \cite{11a}.
Using Eqs. (\ref{33}), (\ref{53})-(\ref{55}), it is easily shown that
\begin{eqnarray}
\label{74}
\left( \frac{\phi}{M_P} \right)^2 & = & \frac{C}{4 \pi}
\int_{D_i}^D \frac{d D' (D'-1)}{D'^3 \left[ \frac{1}{D_0}
-\frac{1}{C} \ln \chi_C - \frac{1}{2C} \ln \pi + \frac{1}{2C}
\psi \left( \frac{D'}{2}+1 \right) \right]}\nonumber\\
& + & \left( \frac{\phi_i}{M_P} \right)^2.
\end{eqnarray}
Using a fine approximation, one can neglect the term including the
Euler's psi function, being in the order of $1/C$, with respect to
$\frac{1}{D_0}$. Therefore, Eq. (\ref{74}) can be written as
\begin{equation}
\label{75}
\left( \frac{\phi}{M_P} \right)^2 =
\frac{ C \left[ \frac{1}{D_i} \left( 1 - \frac{1}{2D_i} \right)
-\frac{1}{D} \left( 1 - \frac{1}{2D} \right) \right]}
{4 \pi \left[ \frac{1}{D_0} - \frac{1}{C} \ln \chi_c -
\frac{1}{2C} \ln \pi \right] }+ \left( \frac{\phi_i}{M_P} \right)^2.
\end{equation}
Substituting $\phi=\phi_f$, $D=D_f$ and the corresponding values of $C$
and $D$ into Eq. (\ref{75}), one can get $\phi_i/M_P=3.491$, see Table
2.
Using Eqs. (\ref{32}) and (\ref{75}), one can find a
relation between $\phi(t)$ and $a(t)$ which is given by
\begin{equation}
\label{76}
\left( \frac{\phi}{M_P} \right)^2 = \frac{ \left( -1 + \frac{1}{D_i}
+\frac{1}{2C} \ln \frac{a}{a_i} \right) \ln \frac{a}{a_i} }
{4 \pi \left[ \frac{1}{D_0} - \frac{1}{C} \ln \chi_c - \frac{1}{2C}
\ln \pi \right] } + \left( \frac{\phi_i}{M_P} \right) ^2.
\end{equation}

\section{Concluding remarks}
\noindent
In this paper, we study a simple model of chaotic inflation by
assuming the space dimension as a dynamical parameter.
For both constant and variable space dimension, at the
end of inflationary epoch, we assume the Universe would have
grown to the size of a ping-pong ball.
This is based on cosmological data
and we describe how to derive it.

In the model, there is a free parameter, called $C$. For different
values of $C$ corresponding to $D_P=4, 10, 25$ and $\infty$,
we calculate the e-folding number.
To obtain the number of e-foldings ${\mathcal N}$, we repeat the
analysis of the inflationary cosmology in three-space.
For $D_P=4, 10, 25$ and $\infty$ we get ${\mathcal N} \sim 70, 99, 116$
and $132$, respectively. The dynamical character of the space dimension
increases the number of e-foldings (for constant three-space, we have
58 e-foldings). Using these e-foldings, we obtain the initial and
final value of  the inflaton field. There
is a small shift in the value of $\phi_i$ and $\phi_f$ to larger values.

Using the number of e-foldings, we also obtain the size of the
Universe at the beginning of inflationary epoch.
For $D_P= 10, 25$ and $\infty$, inflation starts from
an initial size less than the Planck length.
We know that in three-space the initial size of the inflationary epoch
is in the order of the Planck length.
Particularly, for $D_P=25$ and $\infty$ the initial size is less than
the minimum size of the model $\delta$ and also the Planck length $l_P$.
For $D_P=10$, the initial size is bigger than $\delta$ and
smaller than $l_P$.
So, we rule out the cases of $D_P \geq 10$ in the model and
conclude an upper limit for the space dimension at the Planck
length, $D_P < 10$.
Our result for an upper limit for the space dimension at the Planck
length is in agreement with our previous result in \cite{12a}.
Here, we take $D_P=4$ corresponding to $C=1678.8$.

\end{document}